# A Mobile Application for Smart House Remote Control System

Amir Rajabzadeh, Ali Reza Manashty, and Zahra Forootan Jahromi

***Abstract*—At the start of the second decade of 21th century, the time has come to make the Smart Houses a reallity for regular use. The different parts of a Smart House are researched but there are still distances from an applicable system, using the modern technology. In this paper we present an overview of the Smart House subsystems necessary for controlling the house using a mobile application efficiently and securely. The sequence diagram of the mobile application connectiing to the server application and also the use-cases possible are presented. The challenges faced in designing the mobile application and illustrating the updated house top plane view in that application, are discussed and soloutions are adapted for it. Finally the designed mobile application was implemented and the important sections of it were described, such as the interactive house top view map which indicates the status of the devices using predefined icons. The facilities to manage the scheduled tasks and defined rules are also implemented in this mobile application that was developed for use in Windows Mobile platform. This application has the capability of connecting to the main server using GPRS mobile internet and SMS. This system is expected to be an important step towards a unified system structure that can be used efficiently in near future regular houses.

## I. INTRODUCTION

WE are now headed into the second decade of 21th century and we are now vitnessing more and more digital devices all around us in our daily life. They indeed influence in our routine living and we can not even imagine one single day without using them. Mobile phone, PC, TV, audio/video player, air conditioner, fridge, oven, and so on are sample of tens of digital devices we have took them for granted as part of our lives. What more digital devices would come to our home at the end of current decade?

Ease of access and use, is the main purpose of many remote controllers we now use for our devices. Their number is getting bigger and bigger each day, as a new device becomes remotely controllable. Speakers, air conditioners, lights, curtains, garage door, TVs and players are already being remote controlled. Now that every single part of our homes can be controlled remotely, why can't we move toward remotely controlling our homes all in one place?

Smart House is not a new term for science society but is still far more away from people's vision and audition. This is because although recent various works has been done in designing the general overview of the possible remote access approaches for controlling devices [1], or in cases simulating the Smart House itself [2], and designing the main server [3], the design and implementaion of a off-the-shelf Smart House remote control application has been limited to simply the computer applications and just in cases mobile [4] and web applications development [5]. Nowadays people spent a noticable amount of time in transportation, without having access to their PCs or having hard time accessing their labtops; instead, they are constantly using their cell-phones/PDAs. Because of this, we designed and implemented a mobile application that can be connected to a server where other access routes such as web application and local windows application can also meet there.

An important question regarding the problem is whether developing the web application can take the place of mobile phone application, due to availablity of web sites through GPRS and WiMax wireless internet? The answer to this question is that even though we can access our home control system through mobile wireless internet and use current mobile browsers, which are now no less powerful than PC browsers, they can not access GSM messaging systems such as SMS, MMS and so on. In the other hand, simultanious accessing to mobile internet services for viewing a complete webpage, such as GPRS, is still expensive, so there will not be any need for designing the home control schedules and rules [5] online; instead, a temporary connection will do the information updating while using the mobile application offline.

In this paper, we present design of a mobile application for accessing and controlling smart house control systems. We also show an implementation overview using Windows Mobile platforms and C# language and also proposed the general outline of the system.

Dr. Amir Rajabzadeh is with Razi University, Computer Engineering Faculty, Bagh-e-Abrisham, Iran, Kermanshah (Corresponding Author)(Mobile: +989123473189; Tel/Fax: +988314283265; e-mail: rajabzadeh@razi.ac.ir).

Ali Reza Manashty is with Razi University, Computer Engineering Faculty, Bagh-e-Abrisham, Iran, Kermanshah (Mobile:+989355332577; fax: +988318359105; e-mail: a.r.manashty@gmail.com).

Zahra Forootan Jahromi is with Razi University, Computer Engineering Faculty, Bagh-e-Abrisham, Iran, Kermanshah (Mobile: +989177922901;e-mail:zahra.forootan@gmail.com).

## II. ENVIRONMENT OVERVIEW

### A. System Behaviour

The Smart House system usually consists of several devices scattered around the house that are linked together using a wired or wireless network. A computer system acts as a server in a node which controls all the information exchange through the network. The server system must have a device manager (middleware) [3] that assists the main application by connecting all the different device controllers through a single interface or the least interfaces possible. The various types of devices in this house can be divided into three categories:

1. Actuator devices, e.g., alarms, lights, windows and doors
2. Sensor devices, e.g., heat, gas, movement and healthcare
3. Actuator/Sensor devices, e.g., robots, air conditioners

All the devices (whether sensor or actuator) can express their status using their controllers, such as whether they are on or off or the job they are currently involving (e.g. closing the door).

As mentioned before, a device manager, which is part of the server computer, controls all the device controllers and provides methods for retrieving their status and sending commands to them. A device driver might also be needed for more complex devices, because each device might be anything from a lamp to a home robot. The server application has the capability of adding newly installed devices and providing the appropriate controlling methods. The controlling signal and status controlling schema for all these devices is availlable through a general interface. We refer any of these devices, as objects. So the application must be object-independent while the device manager is closely in contact with all these objects through their device drivers and appropriate connections (e.g., cable or wireless ethernet). Using the object-independent interface, we can extend the controlling methods to any further possible ways such as web application, mobile application and telephone line; easily without the need of changing the application codes. The subsystems in this Smart House controlling environment is illustrated is Fig. 1.

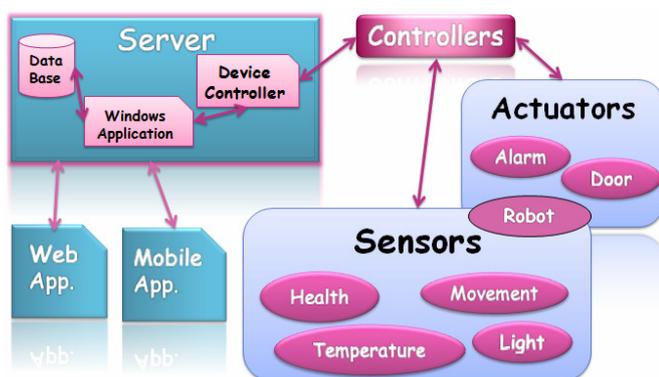

Fig. 1 The subsystems overview of a Smart House controlling system

The database of the system is playing an important role. The connection of the web application and mobile application to the home control system will be through the information in the database.

There are several parts of the system that must be designed and linked together. From the subsystems showed in Fig. 1, we are going to design and implement the mobile application; because, as mentioned in the Introduction section, the other systems had been designed before.

Before desiginig the application itself, we must design the server connection schema of the system, in which the connections of the different parts are modeled.

### B. Server Design

The server is actually the computer system in the smart house that contains the windows application and device manager. The server retrieves the sensor information on regular intervals and updates the database. This interval is different for vital and non-vital devices. For example, elderly and disabled person's health monitoring sensors [6] [7] data must be updated at least every second. The burglary detection system can be updated every few seconds but the temprature and light sensors might be checked within minutes. This gategories help energy-saving schemas to be appliable.

This database also contains information about the devices, scheduled tasks, rules, user access and other policies. Other remote control applications, such as web and mobile application, retrieve sensor and device information from this database and update the scheduled tasks and rules accordingly so each time server updates the database sensor information, it also checkes the changes applied to rules and schedules and perform the actions necessary.

The web application can connect directly to the database, but because of functional restriction, mobile devices can not connect directly to the server database and update the information like regular data connections. So here we face two important issues. The first issue is where to keep the database, so remote application can have continous access to it? The answer to this question is completely related to the web application and security. Web applications can be either hosted in the Smart House server or a host and domain reseller server. Due to security measures, we decided to make the web server and the database all in same place in the home server. This requires a static IP for the house, which is not a problem as security comes first. So the location of the web application is in the Smart Home server and the database is shared between the web and windows application. The second issue is the connection of the mobile application to database for retrieving and updating the information. As mentioned before, current mobile applications may not have the required memory and libraries required to establishing a direct connection to the server, so the only way we can exchange information is through http web servers using GPRS (small data packets can be sent through SMS to the GSM modem attached to server). One possible way to do this task is using web *requests*. Web requests are parameters send to web site using the "?" operator after the webpage name (e.g., www.test.test/login.aspx?user=admin&&pass=123456). After processing the web page requests, it can be identified that a request has been set using the mobile application (using the appropriate web *requests*), and the page *response* will be changed according to the request.

The web response page is reguraly the html content of the website, but using ASP.Net application, before the page can be loaded, according to the web *requests* we can send limited lines of information instead of the whole html page. So we can use this feature to exchange information between the mobile application and a simple aspx page we already developed in our web server.

Now we described the overal server behaviour. From here on, we focus on the mobile application design and implementaion.

### III. DESINGING THE MOBILE APPLICATION

*A. Use-Cases*

The main server computer, which is located in the smart house area, is loaded with the windows application that gives the administrator user a comprehenisve set of options and capabilities. The user can add and manage devices in the application (of course if hardware procedures had been proceeded previously), design the home top view plane using graphical tools and icons, manage user access controls (e.g., define access limits for children), define policies of remote access (e.g., authendicated phone numbers), define rules (conditions to be checked and actions taken if the predefined criteria is met), define scheduled tasks (a task to be done by a specific time (e.g., now)) and check the current status of devices (Fig. 2).

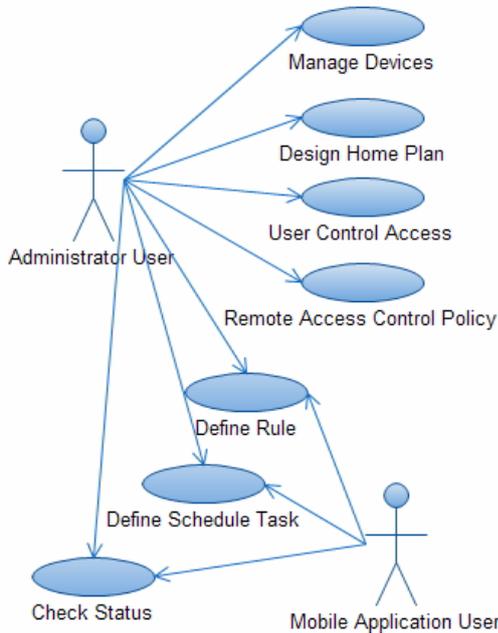

Fig. 2 Use-cases of the Administrator and Mobile user

In the other hand, mobile application user can do the most common and important tasks, but not all the functionalities, as illustrated in Fig. 2. This limitation is mainly because of limits of a mobile application implementation and security factors. For example the mobile application user can not design the home top view plane that graphically shows the current status of the house, but can simply view it from his/her cell phone.

*B. Server connection*

As mentioned earlier, we must connect to the database using a web server. The web application can have most of the capabilities and access levels of the windows application of the server with some restricted regulations. We can allocate some pages of the web application for allowing access from the mobile application that responds to the web requests sent from the mobile http connection.

Because this approach of sending/receiving information is not encrypted, we use encryption algorithms known for both the web and mobile application. For extra security, we use a magic number used for hashing information that expires soon and need to be reconfigured by the web server.

The first attempt to connect to web server with the special code will give the encrypted magic number to the mobile application. Then the user name and password will be sent using a hashing algorithm by the magic number as the salt, as the web *request* parameter. Then the server returns the authentication acknowledge back to the mobile application. Now every request from the server, such as request for updating and checking the status of the devices, must be accompanied by the hashed username and password. After some predefiend time (e.g., 5 minutes), the magic number expires and the server data packets must include newly hashed username and password using the new magic number.

The above sequence is shown in Fig. 3, which aslo shows the sequence of main server application, retrieving information from the devices and database while making the changes necessary. As soon as the mobile application or web application updates the database, the server will check the updated information, time and conditions, so that it will send neccassary commands to the actuators to make the changes applied.

The database contains a table in which icons of the devices and their respective positions will be stored. This information is also transferred to the device in the case of updating the information. This map table information is related but not depended on the main device table (in which detailed device controlling information is stored). In other way, the records and items in the home top view plane can be in the device records list either, but not necessarily all the map items must have full identification record in the device records table. This makes the whole house map items easy designing, but not limiting the selection to controllable devices. Now we must find a suitable way to transfer this map to the mobile device. The problem is the file size of the map that is too much to be transferred easily through the GPRS mobile internet. So we designed a way to create the map in the device instead, using pre-defined icons.

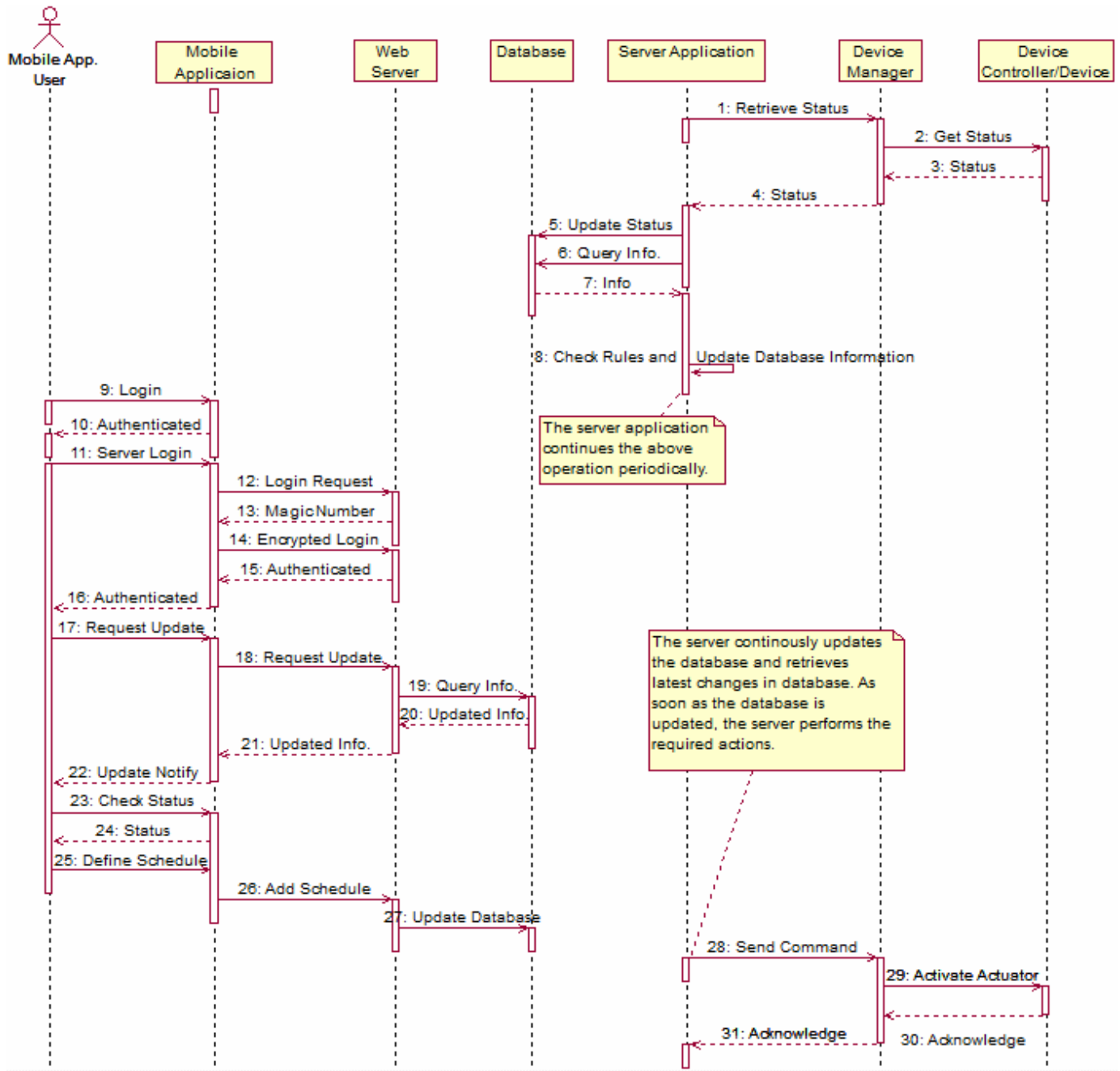

Fig. 3 The sequence diagram of defining a schedule using the mobile application. This UML sequence diagram shows the neccassary steps from logging into mobile application, defining a task and taking the appropriate actions as soon as application server becomes aware of it.

## IV. IMPLEMENTING THE MOBILE APPLICATION

### A. Mobile Map

To transfer the data and create the map in the mobile device, first we must separate the map plane and objects within it. To do the task, as the device requests for updating the information of the devices and the maps, we send the house wall information as arrays of connected lines. To make this thing happen, we actually need a List<List<Points>> (a List of List of Points). The inner list contains points that makes continues lines within the points of the current list, making an open polygon designing possible. The first two points of each inner list will be used to determine the width and color of the lines. Because every point has two integer elements (x, y), two points make 4 integer data integration possible, so one is used for the line width and the other three represents an RGB color value.

For transfering the icons data, we used a record with the fields illustrated in Fig. 4. *OID* field, if not zero, can represent a device in the according device table in the database that can make the device selectable (e.g., for further details view and schedule/rule assignment). Other devices in the map have an *OID* of zero. *Name* and *Position* fields of the device map record are used for displaying purposes. The *IconID* field, regardless of the status of the device in the appropriate device record, indicates the current status of the device/furniture using an icon in the mobile application icon database. The web page map controller is responsible for representing the appropriate *IconID* that best defines the *type* and *current status* of the device. For example, two icons can represent a door in two different statuses of being closed or opened.

| OID | Name | Position | IconID |
|-----|------|----------|--------|

Fig 4 Fields in the record used to transfer home top view plane icons to the mobile application

For new devices that their icons is not available in the mobile application, some extra icons has been considered that makes the other unknown devices into 4 categories that can be recognized by their status easily:
1. On/Off devices (e.g., a lamp)
2. Leveled Devices (e.g., a gas sensor)
3. Appearing/Disappearing devices (e.g., a car or a bike)
4. Opened/Closed Devices (e.g., a door)

The device then repaints the map using the received information from the server, and the icons in its database. First the areas of the house are drawn using the line information defined by the points and the icons are painted just after it. The controllable devices in the map (whose OID is not zero) can be pointed and selected (like the application of the Gator Tech Smart House [8]) to check the status and define schedules for it.

### B. Updating from the server

For implementing the updating procedure, we divide the updating data into two categories. The first part is the devices data table which includes information about the devices, as well as the capabilities and controllable parts of each one. Because this information may be quite large to transfer and the devices and their controllable/sensable features is device dependant, but not state dependant; they can be updated in longer periods than state information. This type of updating is labeled *Update Devices Data* in the main menu of the mobile application.

The second part of the information is the device states and map information. Because these informations are more likely to be updated, and contain less data than the first part, they can be downloaded every time the statuses are being checked. This regular information updating is accessible as *Update Information* button in the main menu form of the mobile application, as well as in *Check Status* and *Home Top Plane View* forms.

### C. Implementaion in Windows Mobile platform

There are several platforms in which it is possible to implement the designed application, such as Java 2 Micro Edition (J2ME), Windows Mobile and Symbian. The J2ME is the most common platforms supported in mobile devices, but it's low level libraries makes it difficult to implement the application in the first place after designing it. So we decided to implement the application, in this stage, using Windows Mobile platform, which can help implementing the application with all the possible features as necessary.

*1) Login Screens*

For increased security, both the mobile application and server connection require a username and password. This was done so that the server address can also be protected from unauthorized viewing. (Fig. 5)

  *a)    User Login*

This screen simply contains the username and password fields for the user to access the application.

  *b)    Server Login*

Just like the simple User Login form, but with an additional textbox field labeled *Server path* that indicates the home web server address that the mobile application must connect to communicate with the server application

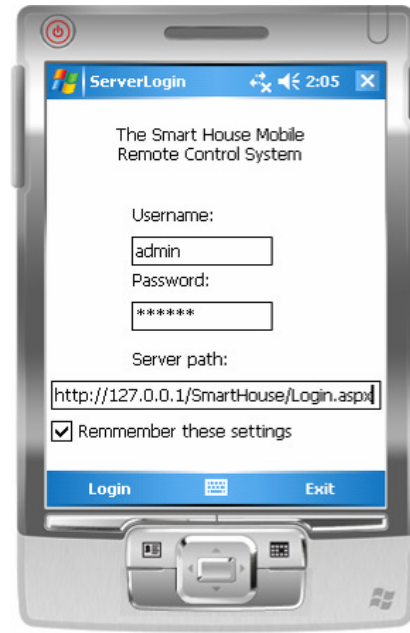

Fig. 5 The server login form of the application. In addition to username and password of the server, the address of the server is also required.

*2) Main Menu*

The main menu form of the application is the form appearing just after logging in. This form contains all the links necessary for different parts of the program (Fig. 6).

*Update Information* and *Update Devices data* was mentioned before. The *Home Top Plane View* will bring up the home top plane map of the house using the most recent updated information.

*Live Camera Streaming* is designed for live streaming the camera devices in the smart house. The *Manage Schedule* and *Manage Rules* items are designed to list and manage the current device scheduled tasks and conditional rules. Scheduled tasks are the tasks which will be done in a specific time (or current moment). Rules are condition and action sets in which the condition are simultaneously checked and as soon as the criteria is met; the appropriate actions will be taken.

The settings option will allow the user to change the server settings such as the address and update rules; as well as changing username/password and other settings.

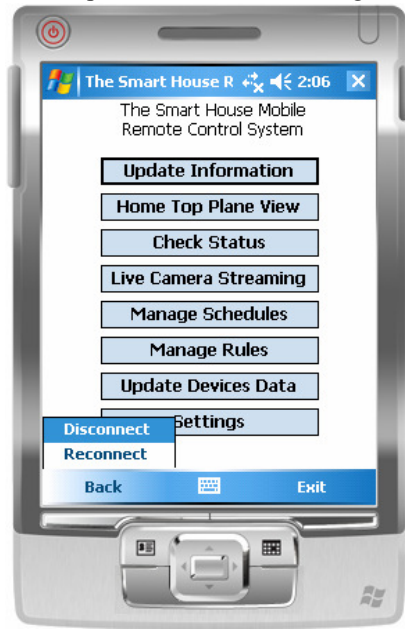

Fig. 6 The main menu form of the mobile application. All different parts of the application can be accessed through this form.

*3) Home top view plane map*

This form shows the home top plane map of the house, designed by the administrative user in the server computer, and downloaded as records of map elements by the mobile application (Fig. 7). If the information is out of date, it prompts the user to update the information. The user can also select the devices here and be promped whether he/she wants to check its status or add a schedule for it and then will be redirected to the appropriate form. The icons in this map also reperesnts the current status of the device as indicated by the *IconID* field of the device map record retrieved from the server.

*4) Managing Schedule*

In this form, the updated schedules will be listed and each of them can be modified and be enabled/disabled (Fig. 8). Each task has a name and will be applied on a device by the corresponding device action. The time of the scheduled task can be set to now or a specific time in the day. Some criteria can also be set, so that only if the desired conditions are met, the task will be taken place. When some task will be done, it will be disabled. These data can be sent using SMS either.

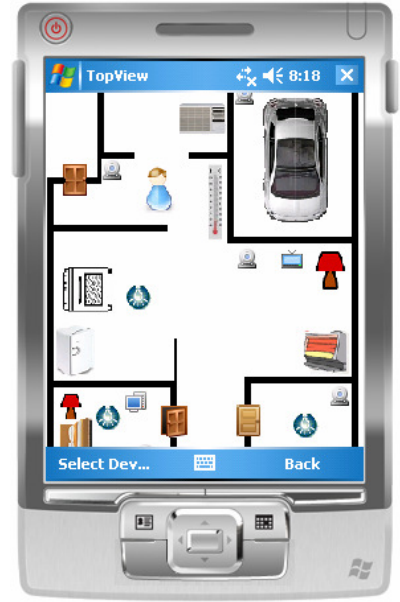

Fig. 7 Home top view plane of the house with selectable items.

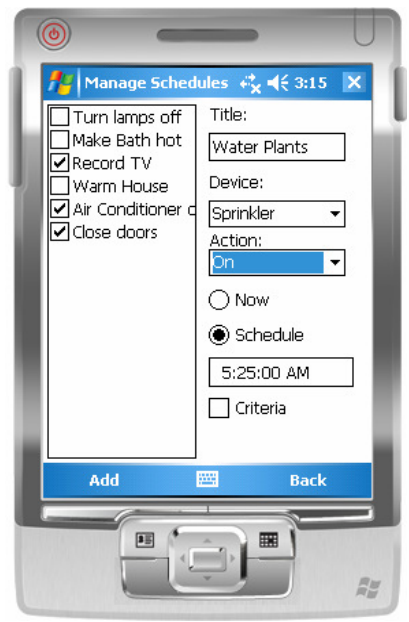

Fig. 8 The schedules management form of the mobile application.

## V. FUTURE WORKS

As mentioned earlier, the works on a complete and comprehensive Smart House that can work with all possible home appliances and can be controlled by all means in a effective way, are all scattered around and researched independently. Only in some projects, some parts of a real smart house put into practice (e.g., The Gator Tech Smart House [8] and [9]). Now that we designed a server and mobile application needed for controlling the smart house remotely, we must continue the work on completing the whole server and applications, and move through commercial manufacturing of such houses so that all these efforts on designing the smart house can come to reality.

## VI. CONCLUSION

In this paper, we presented an overview of a smart house control system server along with the way the devices are managed in the server. After discussing the possible security issues of developing a server for communication to mobile application, we proposed a web server for the mobile application to communicate to it using GPRS. We presented the communication sequence through the web server in a UML sequence diagram and described the use-cases of both the server and mobile application. We finally explained the design of the mobile application and the data records needed for transferring the data and home top view plane from the server to mobile application in an efficient way. We finally described the main parts of the implementaion of this smart house remote control mobile application in the Windows Mobile platform.